\def\be{\begin{equation}}
\def\ee{\end{equation}}
\def\ba{\begin{array}}
\def\p{\prime}
\def\ea{\end{array}}

\def\Rb{{I\!\! R}}

\documentstyle[12pt]{article}
\begin{document}
\parskip=4pt
\parindent=18pt
\baselineskip=24pt
\setcounter{page}{1}
\centerline{\Large\bf A Remark on One-Dimensional Many-Body}
\vspace{1ex}
\centerline{\Large\bf Problems with Point Interactions}
\vspace{6ex}
\parindent=18pt
\parskip=6pt
\begin{center}
{\large  Sergio Albeverio}\footnote {SFB 256; SFB 237; BiBoS;
CERFIM (Locarno); Acc.Arch., USI (Mendrisio)},
{\large Ludwik D{\c a}browski}\footnote{
SISSA, I-34014 Trieste, Italy} and
{\large  Shao-Ming Fei}\footnote{Institute of Physics,
Chinese Academy of Science, Beijing.}

\end{center}
\begin{center}
Institut f\"ur Angewandte Mathematik, Universit\"at Bonn, D-53115 Bonn\\
and\\
Fakult\"at f\"ur Mathematik, Ruhr-Universit\"at Bochum, D-44780 Bochum
\end{center}
\vskip 1 true cm

\begin{center}
\begin{minipage}{5in}
\vspace{3ex}
\centerline{\large Abstract}
\vspace{4ex}
The integrability of one dimensional quantum mechanical many-body
problems with general contact interactions is extensively studied. 
It is shown that besides the pure (repulsive or attractive) 
$\delta$-function interaction there is another singular 
point interactions which gives rise to a new one-parameter family
of integrable quantum mechanical many-body systems. 
The bound states and scattering matrices 
are calculated for both bosonic and fermionic statistics.

\end{minipage}
\end{center}


\newpage

Quantum mechanical solvable models describing a particle moving in a
local singular potential concentrated at one or a discrete number of points
have been extensively discussed in the literature, see e.g.
\cite{agh-kh,gaudin,AKbook} and references therein.
One dimensional problems with contact interactions at, say, the origin
($x=0$) can be characterized by separated or nonseparated
boundary conditions imposed on the (scalar) wave function $\varphi$ at
$x=0$. The classification of one dimensional point interactions
in terms of singular perturbations is given in \cite{kurasov}.
In the present paper we are interested in many-body problems with
pairwise interactions given by such singular potentials.
The first model of this type
with the pairwise interactions determined by $\delta$-functions
was suggested and investigated
in \cite{mcguire}. Intensive studies of this model applied to statistical
mechanics (particles having boson or fermion statistics)
are given in \cite{y,y1} (these also leads to the well known 
Yang-Baxter equations).

Nonseparated boundary conditions correspond
to the cases where the perturbed operator is equal to
the orthogonal sum of two self-adjoint operators in $ L_{2}
(-\infty,0] $ and $L_{2} [0,\infty)$.
The family of point interactions for the one dimensional
Schr\"odinger operator $ - \frac{d^2}{dx^2}$
can be described by unitary $ 2 \times  2 $ matrices
via von Neumann formulas for self-adjoint extensions
of symmetric operators, since the second derivative
operator restricted to the domain $ C_{0}^\infty ({\bf R}
\setminus \{ 0 \} ) $ has deficiency indices $ (2,2)$.
The boundary conditions describing the self-adjoint extensions
have the following form
\begin{equation} \label{bound}
\left( \begin{array}{c}
\varphi\\
\varphi '\end{array} \right)_{0^+}
= e^{i\theta} \left(
\begin{array}{cc}
a & b \\
c & d \end{array} \right)
\left( \begin{array}{c}
\varphi\\
\varphi '\end{array} \right)_{0^-},
\end{equation}
where
\be\label{abcd}
ad-bc = 1,~~~~\theta, a,b,c,d \in \Rb.
\ee
$\varphi(x)$ is the scalar wave function of two spinless particles with
relative coordinate $x$. (\ref{bound}) also describes two particles
with spin $s$ but without any spin coupling between the particles when they
meet (i.e. for $x=0$), in this case $\varphi$ represents any one of the
components of the wave function. The values $\theta = b=0$, $a=d=1$ in
(\ref{bound})
correspond to the case of a positive (resp. negative) $\delta$-function
potential for $c>0$ (resp. $c<0$). For general $a,b,c$ and $d$, the
properties of the corresponding Hamiltonian systems have been studied in
detail, see e.g. \cite{kurasov,ch,abd}.

The separated boundary conditions are described by
\be\label{bounds}
\varphi^\prime(0_+) = h^+ \varphi (0_+)~, ~~~
\varphi^\prime(0_-) = h^- \varphi (0_-),
\ee
where $h^{\pm} \in \Rb \cup \{ \infty\}$.
$ h^+ = \infty$ or $ h^- = \infty$
correspond to Dirichlet boundary conditions and
$ h^+ = 0$ ~or~ $ h^- = 0$ correspond to 
Neumann boundary conditions.
In this case it is impossible to express the perturbed operator
as the orthogonal sum of two self-adjoint operators in $ L_{2}
(-\infty,0] $ and $L_{2} [0,\infty)$.

In the following we study the
integrability of one dimensional systems of $N$-identical particles
with general contact
interactions described by the boundary conditions
(\ref{bound}) or (\ref{bounds})
that are imposed on the relative coordinates of the particles.
We first consider the case of two particles ($N=2$) with
coordinates $x_1$, $x_2$ and momenta $k_1$, $k_2$ respectively. 
Each particle has $n$-`spin' states designated by $s_1$ and $s_2$,
$1\leq s_i\leq n$. For $x_1\neq x_2$, these two particles are free. The
wave functions $\varphi$ are symmetric (resp. antisymmetric) with respect
to the interchange $(x_1,s_1)\leftrightarrow(x_2,s_2)$ for bosons (resp.
fermions). In the region $x_1<x_2$, from the Bethe ansatz the
wave function is of the form,
\be\label{w1}
\varphi=\alpha_{12}e^{i(k_1x_1+k_2x_2)}+\alpha_{21}e^{i(k_2x_1+k_1x_2)},
\ee
where $\alpha_{12}$ and $\alpha_{21}$ are $n^2\times 1$ column matrices.
In the region $x_1>x_2$,
\be\label{w2}
\varphi=(P^{12}\alpha_{12})e^{i(k_1x_2+k_2x_1)}
+(P^{12}\alpha_{21})e^{i(k_2x_2+k_1x_1)},
\ee
where according to the symmetry or antisymmetry conditions,
$P^{12}=p^{12}$ for bosons and $P^{12}=-p^{12}$ for fermions, $p^{12}$
being the operator on the $n^2\times 1$ column that interchanges
$s_1\leftrightarrow s_2$. 

Let $k_{12} = (k_1 -k_2)/2$. In the center of mass
coordinate $X=(x_1+x_2)/2$ and the relative coordinate
$x=x_2-x_1$, we get, by substituting (\ref{w1}) and (\ref{w2}) into the
boundary conditions at $x=0$, 
\be\label{a1}
\left\{
\begin{array}{l} 
\alpha_{12}+\alpha_{21}
=e^{i\theta}aP^{12}(\alpha_{12}+\alpha_{21})+
ie^{i\theta}bk_{12}P^{12}(\alpha_{12}-\alpha_{21}),\\
ik_{12}(\alpha_{21}-\alpha_{12})
=
e^{i\theta}cP^{12}(\alpha_{12}+\alpha_{21})+ie^{i\theta}dk_{12}P^{12}
(\alpha_{12}-\alpha_{21})
\end{array}\right.
\ee
for boundary condition (\ref{bound}), and
\be\label{a2}
\left\{
\begin{array}{l} 
ik_{12}(\alpha_{21}-\alpha_{12})
=
h_+ (\alpha_{12}+\alpha_{21})~,\\
ik_{12}P^{12}(\alpha_{12}-\alpha_{21})
=
h_- P^{12}(\alpha_{12}+\alpha_{21})
\end{array}\right.
\ee
for boundary condition (\ref{bounds}) respectively.

Eliminating the term $P^{12}\alpha_{12}$ from (\ref{a1})
we obtain the relation 
\be\label{2112}
\alpha_{21} = Y_{21}^{12} \alpha_{12}~,
\ee
where
\be\label{a21a12}
Y_{21}^{12}
=\frac{
2ie^{i\theta}k_{12}P^{12}+ik_{12}(a-d)+(k_{12})^2b+c
}{
ik_{12}(a+d) + (k_{12})^2b-c}.
\ee

We remark that the system (\ref{a2}) is contradictory unless
\be\label{hh}
h_+ = - h_-\doteq h \in \Rb \cup \{ \infty\}.
\ee
In this case it also leads to equation (\ref{2112}) with
\be\label{a21a12s}
Y_{21}^{12}
=\frac{ik_{12} + h}{ik_{12} - h} ~.
\ee

For $N\geq 3$ and $x_1<x_2<...<x_N$, the wave function is given by
\be\label{psi}
\psi=\alpha_{12...N}e^{i(k_1x_1+k_2x_2+...+k_Nx_N)}
+\alpha_{21...N}e^{i(k_2x_1+k_1x_2+...+k_Nx_N)}+(N!-2)~other~terms.
\ee
The columns $\alpha$ have $n^N\times 1$ dimensions. The wave functions
in the other regions are determined from (\ref{psi}) by the requirement of
symmetry (for bosons) or antisymmetry (for fermions).
Along any plane $x_i=x_{i+1}$, $i\in 1,2,...,N-1$, from similar
considerations as above we have
\be\label{a1n}
\alpha_{l_1l_2...l_il_{i+1}...l_N}=Y_{l_{i+1}l_i}^{ii+1}
\alpha_{l_1l_2...l_{i+1}l_i...l_N},
\ee
where
\be\label{y}
Y_{l_{i+1}l_i}^{ii+1}=
\frac{2ie^{i\theta}k_{l_il_{i+1}}P^{ii+1}
+ik_{l_il_{i+1}}(a-d) + (k_{l_il_{i+1}})^2 b+c}
{ik_{l_il_{i+1}}(a+d)+(k_{l_il_{i+1}})^2 b-c}
\ee
for nonseparated boundary condition and 
\be\label{ys}
Y_{l_{i+1}l_i}^{ii+1}=
\frac{ik_{l_il_{i+1}} + h}{ik_{l_il_{i+1}} - h} 
\ee
for separated boundary condition.
Here $k_{l_il_{i+1}}=(k_{l_i}-k_{l_{i+1}})/2$ play the role of spectral
parameters.
$P^{ii+1}=p^{ii+1}$ for bosons and $P^{ii+1}=-p^{ii+1}$ for fermions,
with $p^{ii+1}$ the operator on the $n^N\times 1$ column
that interchanges $s_i\leftrightarrow s_{i+1}$.

For consistency $Y$ must satisfy the Yang-Baxter equation with
spectral parameter \cite{y,ma}, i.e.,
$$
Y^{m,m+1}_{ij}Y^{m+1,m+2}_{kj}Y^{m,m+1}_{ki}
=Y^{m+1,m+2}_{ki}Y^{m,m+1}_{kj}Y^{m+1,m+2}_{ij},
$$
or
\be\label{ybe1}
Y^{mr}_{ij}Y^{rs}_{kj}Y^{mr}_{ki}
=Y^{rs}_{ki}Y^{mr}_{kj}Y^{rs}_{ij}
\ee
if $m,r,s$ are all unequal, and
\be\label{ybe2}
Y^{mr}_{ij}Y^{mr}_{ji}=1,~~~~~~
Y^{mr}_{ij}Y^{sq}_{kl}=Y^{sq}_{kl}Y^{mr}_{ij}
\ee
if $m,r,s,q$ are all unequal.

The operators $Y$ given by (\ref{y}) satisfy the relation (\ref{ybe2}) 
for all $\theta ,a,b,c,d$. However the relations (\ref{ybe1}) are
satisfied only when $\theta =0$, $a=d$ and $b=0$, that
is, according to the constraint (\ref{abcd}),
$\theta =0$, $a=d=\pm 1$, $b=0$, $c$ arbitrary.
The case $a=d=1$, $\theta =b=0$ corresponds to the usual $\delta$-function
interactions, which has been investigated in \cite{y,y1}.
The case $a=d=-1$, $\theta =b=0$, 
which we shall refer to as `anti-$\delta$' interaction,
is related to another singular interactions between any pair of particles
(for $a=d=-1$ and $\theta =b=c=0$ see \cite{kurasov,ch}). Associated with the
separated boundary condition, the operators
$Y$ given by (\ref{ys}) satisfy both the
relations (\ref{ybe2}) and (\ref{ybe1}) for arbitrary $h$.

We have thus found that with respect to $N$-particle (either boson or
fermion) problems, altogether there are three integrable one parameter
families
with contact interactions of type $\delta$, anti-$\delta$ and separated one,
described respectively by one of the following conditions on the wave
function along the plane $x_i=x_j$ for any pair of particles
with coordinates $x_i$ and $x_j$, 
\be\label{b0}
\varphi(0_+)=+\varphi(0_-),~~~\varphi^\prime(0_+)=c\varphi(0_-)+\varphi^\prime(0_-)~, ~
c\in \Rb ~;
\ee
\be\label{b}
\varphi(0_+)=-\varphi(0_-),~~~\varphi^\prime(0_+)=c\varphi(0_-)-\varphi^\prime(0_-)~, ~
c\in \Rb ~;
\ee
\be\label{bs}
\varphi^\prime (0_+)= h \varphi(0_+),~~~\varphi^\prime(0_-)= -h \varphi(0_-)~, ~
h\in \Rb \cup \{ \infty\}~.
\ee
The wave functions are given by (\ref{psi}) with the
$\alpha$'s determined by (\ref{a1n}) and initial conditions. The
operators $Y$ in (\ref{a1n}) are given respectively by
\be\label{y0}
Y_{l_{i+1}l_i}^{ii+1}= \frac{i(k_{l_i}-k_{l_{i+1}})P^{ii+1} +c}
{i(k_{l_i}-k_{l_{i+1}}) - c} ~;
\ee
\be\label{y1}
Y_{l_{i+1}l_i}^{ii+1}=-\frac{i(k_{l_i}-k_{l_{i+1}})P^{ii+1} +c}
{i(k_{l_i}-k_{l_{i+1}})+c} ~;
\ee
and
\be\label{y1s}
Y_{l_{i+1}l_i}^{ii+1}
=\frac{i(k_{l_i}-k_{l_{i+1}}) +2 h}{i(k_{l_i}-k_{l_{i+1}}) - 2 h} ~.
\ee

Nevertheless, from (\ref{y0}) and (\ref{y1}) we see that
if we simultaneously change $c \to -c$ and $P^{ii+1} \to -P^{ii+1}$,
these two formulas are interchanged.
There is a sort of duality between bosons (resp. fermions) with 
$\delta$-interaction of strength $c$ and fermions (resp. bosons)
with anti-$\delta$ interaction of strength $-c$.
It can be checked that under the ``kink type'' gauge transformation
${\cal U} = \prod_{i>j}{\rm ~sgn} (x_i-x_j)$,
the N-boson (resp. fermion) $\delta$-type contact interaction 
goes over to the N-fermion (resp. boson) anti-$\delta$ interaction.
Therefore these two situations are in fact unitarily equivalent
under a gauge transformation ${\cal U}$ that is
non-smooth and does not factorize through one particle Hilbert spaces.

The integrable system related to the case (\ref{y1s})
is not unitarily equivalent to either the $\delta$ or anti-$\delta$ cases. 
In fact their spectra are different (see the bound states below).
In the following we study further the one dimensional integrable
$N$-particle systems associated with (\ref{y1s}).

When $h<0$, there exist bound states.
For $N=2$, the space part of the orthogonal basis (labeled by $\pm$)
in the doubly degenerate bound state subspace has the form, 
in the relative coordinate $x=x_2-x_1$,
\be\label{bpsi2s}
\psi_{2,\pm}=
( \theta (x)\pm \theta(-x) )e^{h\vert x\vert}.
\ee
The eigenvalue corresponding to the bound states (\ref{bpsi2s}) is $-h^2$. 
By generalization we get the $2^{N(N-1)/2}$ bound states for
$N$-particle system
\be\label{bpsins}
\psi_{N,\underline{\epsilon}}=
\alpha_{\underline{\epsilon}}
\prod_{k>l} (\theta (x_k-x_l) +\epsilon_{kl}\theta (x_l-x_k))
e^{h\sum_{i>j} \vert x_i-x_j\vert },
\ee
where $\alpha_{\underline{\epsilon}}$ is the spin wave function and 
$\underline{\epsilon} \equiv \{ \epsilon_{kl}~:~k>l \}$; $\epsilon_{kl}=\pm$,
labels the $2^{N(N-1)/2}$-fold degeneracy.

It can be checked that $\psi_{N,\underline{\epsilon}}$ 
satisfies the boundary condition (\ref{bs})
at $x_i=x_j$ for any $i\neq j\in 1,...,N$.
The spin wave
function $\alpha$ here satisfies $P^{ij}\alpha=\epsilon_{ij}\alpha$ 
for any $i\neq j$, that is, $p^{ij}\alpha=\epsilon_{ij}\alpha$ 
for bosons and $p^{ij}\alpha=-\epsilon_{ij}\alpha$ for fermions.
$\psi_{N,\underline{\epsilon}}$ is of the form (\ref{psi}) in each region. 
For instance comparing $\psi_{N,\underline{\epsilon}}$ with (\ref{psi}) 
in the region $x_1<x_2...<x_N$, we get
\be\label{ks}
k_1=ih(N-1),~k_2=k_1-2ih,~k_3=k_2-2ih,...,k_N=-k_1.
\ee
The energy of the bound state $\psi_{N,\underline{\epsilon}}$ is
\be\label{es}
E=-\frac{h^2}{3}N(N^2-1)~.
\ee

The scattering matrix can readily be discussed. For real
$k_1<k_2<...k_N$, in each coordinate region such as $x_1<x_2<...x_N$,
the following term in (\ref{psi}) is an outgoing wave
\be\label{out}
\psi_{out}=\alpha_{12...N}e^{k_1x_1+...+k_Nx_N}.
\ee
An incoming wave with the same exponential as (\ref{out}) is given by
\be\label{in}
\psi_{in}=[P^{1N}P^{2(N-1)}...]\alpha_{N(N-1)...1}e^{k_Nx_N+...+k_1x_1}
\ee
in the region $x_N<x_{N-1}<...<x_1$. The scattering matrix is defined by
$\psi_{out}=S\psi_{in}$. From (\ref{a1n}) we have
$$
\ba{l}
\alpha_{12...N}=[Y_{21}^{12}Y_{31}^{23}...Y_{N1}^{(N-1)N}]\alpha_{2...N1}
=...\\[4mm]
=[Y_{21}^{12}Y_{31}^{23}...Y_{N1}^{(N-1)N}]
[Y_{32}^{12}Y_{42}^{23}...Y_{N2}^{(N-2)(N-1)}]
...[Y_{N(N-1)}^{12}]\alpha_{N(N-1)...1}
\equiv S^\prime \alpha_{N(N-1)...1},
\ea
$$
where $Y_{l_{i+1}l_i}^{ii+1}$ is given by (\ref{y1s}).
Therefore
$$
S=S^\prime P^{N1}P^{(N-1)2}...P^{1N}
=S^\prime [P^{12}][P^{23}P^{12}][P^{34}P^{23}P^{12}]...
[P^{(N-1)N}...P^{12}].
$$
Defining
\be\label{xij}
X_{ij}=Y^{ij}_{ij}P^{ij}
\ee
we obtain
\be\label{s}
S=[X_{21}X_{31}...X_{N1}][X_{32}X_{42}...X_{N2}]...[X_{N(N-1)}].
\ee
The scattering matrix $S$ is unitary and
symmetric due to the time reversal invariance of the interactions.
$<s_1^\p s_2^\p...s_N^\p\vert S\vert s_1s_2...s_N>$ stands for the $S$
matrix element of the process from the state $(k_1s_1,k_2s_2,...,k_Ns_N)$ to
the state $(k_1s_1^\p,k_2s_2^\p,...,k_Ns_N^\p)$.
The momenta (\ref{ks}) are imaginary for bound
states. The scattering of clusters (bound states) 
can be discussed in a similar way as in \cite{y1}. 
For instance for the scattering of a bound state of two
particles ($x_1<x_2$) on a bound state of three particles
($x_3<x_4<x_5$), the scattering matrix is $S=[X_{32}X_{42}X_{52}]
[X_{31}X_{41}X_{51}]$.

We have extensively investigated the integrability of
one dimensional quantum mechanical many-body
problems with general contact interactions.
Besides the repulsive or attractive
$\delta$ and anti-$\delta$ function interactions,
there is another integrable one parameter
families associated with separated boundary conditions.
From our calculations it is clear that these are all 
the integrable systems
for one dimensional quantum identical many-particle
models (of fermionic or bosonic statistics) with
contact interactions. Here the possible contact coupling of the spins of
two particles are not taken into account. A further study along this
direction would possibly give rise to more interesting integrable quantum 
many-body systems.

\vspace{2.5ex}
\noindent
ACKNOWLEDGEMENTS: We would like to thank P. Kulish, P. Kurasov and
V. Rittenberg for helpful comments.


\vspace{2.5ex}
\begin{thebibliography}{20}
\bibitem{agh-kh}
S. Albeverio, F. Gesztesy, R. H\o egh-Krohn and H. Holden, {\it
Solvable Models in Quantum Mechanics}, New York: Springer, 1988.

\bibitem{gaudin}
M. Gaudin, {\it La fonction d'onde de Bethe}, Masson, 1983.

\bibitem{AKbook}
S. Albeverio and R. Kurasov, {\it Singular perturbations of differential
operators and solvable Schr\"odinger type operators},
Cambridge Univ. Press., to appear in 1999.

\bibitem{kurasov}
P. Kurasov, {\it Distribution theory for discontinuous test functions and
differential operators with generalized coefficients}, J. Math. Analy. Appl. 
{\bf 201}(1996)297-323.

\bibitem{mcguire} J.B. McGuire, Study of exactly soluble
one--dimensional N--body problems,
{\it J.Math.Phys.}, {\bf 5} (1964), 622--636.\\
J.B. McGuire, Interacting fermions in one
dimension.I. Repulsive potential,
{\it J.Math.Phys.}, {\bf 6} (1965), 432--439.\\
J.B. McGuire, Interacting fermions in one
dimension.II. Attractive potential,
{\it J.Math.Phys.}, {\bf 7} (1966), 123--132.\\
J.B. McGuire and C.A. Hurst,
The scattering of three impenetrable particles in one dimension,
{\it J.Math.Phys.}, {\bf 13} (1972), 1595--1607.\\
J.B. McGuire and C.A.Hurst,
Three interacting particles in one dimension:
an algebraic approach,
{\it J.Math.Phys.}, {\bf 29} (1988), 155--168.

\bibitem{y}
C.N. Yang, {\it Some exact results for the many-body problem
in one dimension with repulsive delta-function interaction}, Phys. Rev.
Lett. {\bf 19}(1967)1312-1315.\\
C.N. Yang, {\it $S$ matrix for the one-dimensional $N$-body problem with
repulsive $\delta $-function interaction}, Phys. Rev. 
{\bf 168}(1968)1920-1923.

\bibitem{y1}
C.H. Gu and C.N. Yang, {\it A one-dimensional $N$ Fermion
problem with factorized $S$ matrix}, Commun. Math. Phys. {\bf 122}
(1989)105-116.

\bibitem{ch}
P. Chernoff and R. Hughes, {\it A new class of point interactions in one
dimension}, J. Func. Anal. {\bf 111}(1993)97-117.

\bibitem{abd}
S. Albeverio, Z. Brze\'{z}niak and L D\c{a}browski, {\it
Time-dependent propagator with point interaction}, 
J. Phys. A{\bf 27}(1994)4933-4943.

\bibitem{ma}
Z.Q. Ma, {\it Yang-Baxter Equation and Quantum Enveloping Algebras},
World Scientific, 1993.\\
V. Chari and A. Pressley, {\it A Guide to Quantum Groups}, Cambridge
University Press, 1994.\\
C. Kassel, {\it Quantum Groups}, Springer-Verlag, New-York, 1995.\\
S. Majid, {\it Foundations of Quantum Group Theory}, Cambridge
University Press, 1995.\\
K. Schm\"udgen, {\it Quantum Groups and Their Representations},
Springer, 1997.

\end{thebibliography}
\end{document}